\documentclass[%
aip,
amsmath,amssymb,
reprint,%
]{revtex4-1}

\usepackage{graphicx}% Include figure files
\usepackage{dcolumn}% Align table columns on decimal point
\usepackage{bm}% bold math
\usepackage{makeidx}
\usepackage{afterpage}
\usepackage{braket}		%Dirac notation
\usepackage{pgfplots}   %for matlab2tikz graphics handling
\pgfplotsset{compat=newest}
\usetikzlibrary{external}
\usepackage{xcolor}
\usepackage{placeins}
\usepackage{siunitx}
\usepackage{placeins}
\usepackage{mathtools}
\DeclarePairedDelimiter{\abs}{\lvert}{\rvert}

\begin{document}

\title{Broadband single-mode planar waveguides in monolithic 4H-SiC}% Force line breaks with \\

\author{Tom Bosma}
\affiliation{Zernike Institute for Advanced Materials, University of Groningen, NL-9474AG Groningen, the Netherlands}
\author{Joop Hendriks}
\affiliation{Zernike Institute for Advanced Materials, University of Groningen, NL-9474AG Groningen, the Netherlands}
\author{Misagh Ghezellou}
\affiliation{Link\"{o}ping University, Department of Physics, Chemistry and Biology, S-581 83 Link\"{o}ping, Sweden}
\author{Nguyen T. Son}
\affiliation{Link\"{o}ping University, Department of Physics, Chemistry and Biology, S-581 83 Link\"{o}ping, Sweden}
\author{Jawad Ul-Hassan}
\affiliation{Link\"{o}ping University, Department of Physics, Chemistry and Biology, S-581 83 Link\"{o}ping, Sweden}
\author{Caspar H. van der Wal}
\affiliation{Zernike Institute for Advanced Materials, University of Groningen, NL-9474AG Groningen, the Netherlands}

%%%%%%%%%%%%%%%%%%%%%%%%%%%%%%%%%%%%%%%%%%%%%%%%%%%%%%%%%%%%%%%%%%%%%%%%%%%%%%%%%%%%%%%%%%%%%%%%%%%%%%%%%%%%%%%%
%%%%%%%%%%%%%%%%%%%%%%%%%%%%%%%%%%%%%%%%%%%%%%%%%%%%%%%%%%%%%%%%%%%%%%%%%%%%%%%%%%%%%%%%%%%%%%%%%%%%%%%%%%%%%%%%
%%%%%%%%%%%%%%%%%%%%%%%%%%%%%%%%%%%%%%%%%%%%%%%%%%%%%%%%%%%%%%%%%%%%%%%%%%%%%%%%%%%%%%%%%%%%%%%%%%%%%%%%%%%%%%%%

\date{\today}% It is always \today, today,
             %  but any date may be explicitly specified

\begin{abstract}
	Color-center defects in silicon carbide promise opto-electronic quantum applications in several fields, such as computing, sensing and communication. In order to scale down and combine these functionalities with the existing silicon device platforms, it is crucial to consider SiC integrated optics. In recent years many examples of SiC photonic platforms have been shown, like photonic crystal cavities, film-on-insulator waveguides and micro-ring resonators. However, all these examples rely on separating thin films of SiC from substrate wafers. This introduces significant surface roughness, strain and defects in the material, which greatly affects the homogeneity of the optical properties of color centers. Here we present and test a method for fabricating monolithic single-crystal integrated-photonic devices in SiC: tuning optical properties via charge carrier concentration. We fabricated monolithic SiC \mbox{n-i-n} and \mbox{p-i-n} junctions where the intrinsic layer acts as waveguide core, and demonstrate the waveguide functionality for these samples. The propagation losses are below 14~dB/cm. These waveguide types allow for addressing color-centers over a broad wavelength range with low strain-induced inhomogeneity of the optical-transition frequencies. Furthermore, we expect that our findings open the road to fabricating waveguides and devices based on \mbox{p-i-n} junctions, which will allow for integrated electrostatic and radio frequency (RF) control together with high-intensity optical control of defects in silicon carbide.		

\end{abstract}

\maketitle
%%%%%%%%%%%%%%%%%%%%%%%%%%%%%%%%%%%%%%%%
%%%%%%% INTRODUCTION
%%%%%%%%%%%%%%%%%%%%%%%%%%%%%%%%%%%%%%%%

In recent years silicon carbide has gained interest for quantum technology applications in fields like communication \cite{koehl2011,zwier2015,widmann2019, Majety2021} and (bio)sensing \cite{saddow2011,simin2015}. It was found that color centers in SiC have favorable properties, such as long-lived spin states and possibilities for operation at telecom wavelengths \cite{bosma2018, Spindlberger2019a, Wolfowicz2020}. For scaling down potential opto-electronic quantum applications, integrated photonics are paramount. In silicon carbide it would be simple to combine such architectures with existing silicon and silicon-carbide device platforms.

Though many examples exist of amorphous SiC waveguides \cite{pandraud2007,zheng2019,zhang2020}, a high-quality single-crystal material is required in order to get reliable spin-active color centers with predetermined properties \cite{bosma2018}. Crystalline waveguides and photonic devices have been fabricated for various SiC polytypes such as photonic crystal cavities with arrays of cylindrical holes in 3C-SiC \cite{calusine2014}, 4H-SiC film-on-insulator waveguides \cite{lukin2020} and micro-ring resonators \cite{martini2017,zheng2019}.

Inspired by recent studies on color-center defects in silicon carbide, showing combined electrical and optical control of color-center defects in silicon carbide for quantum technologies in \mbox{p-i-n} junctions \cite{falk2014,widmann2018,widmann2019,anderson2019}, we report here on developing another method for fabricating monolithic SiC waveguides: engineering a lower value for the index of refraction $n$ for cladding layers via a high concentration of free charge carriers \cite{hunsperger2009}. The main reason for this choice is that for technologies involving color centers, especially ensemble based, high material purity and low strain-induced inhomogeneity are vital. This grade of material quality, however, can only be achieved when growing 4H-SiC device layers on substrates of the same material. The layers are difficult to separate without loss of homogeneity \cite{zheng2019}.

%-------------------------------------------------------------------
%--------------------- Overview ------------------------------------
%-------------------------------------------------------------------

In this work we show how junctions from layers with alternating doping concentration can be engineered into single-mode optical waveguides. We fabricated \mbox{n-i-n} and \mbox{p-i-n} junction planar structures in 4H-SiC as a proof of concept. These devices can be engineered to confine any preferred number of waveguide modes. Notably, the number of allowed modes, as well as their field distribution, is independent of wavelength for a very wide range, as long as the associated optical frequency is above the plasma frequency. We demonstrate these broadband properties (for 700 to 1290~nm light) of the \mbox{n-i-n} single-mode waveguides with a core thickness of $4~\mu$m and cladding doping concentration up to $10^{19}~\text{cm}^{-3}$, and we present consistent results for \mbox{p-i-n} waveguides. Our findings indicate that it is possible to have electrostatic and RF control for SiC color centers in the core layer of \mbox{n-i-n} and \mbox{p-i-n} structures by the (AC) Stark effect \cite{falk2014,casas2017}, while having high-intensity optical control via the waveguide at the same time. For SiC mature semiconductor processing techniques are available, such that  device structures with one-dimensional photonic channels can be developed.

	%-------------------------------------------------------------------
%--------------------- Samples and predictions ---------------------
%-------------------------------------------------------------------
\section{Tuning the refractive index via carrier-concentration}
According to the Drude model, a plasma of free charge carriers reduces the refractive index \cite{bond1963,bennett1990,fox2010} for optical frequencies above the plasma frequency. They screen a material from the optical electric field, thereby lowering the net polarization and thus the index of refraction $n$. As explained in Appendix~\ref{app:drude}, the influence of free carriers on the refractive index leads to a shift
\begin{equation}\label{eq:refractiveindex}
\Delta n = -\frac{Ne^2\lambda_0^2}{2\varepsilon_0 n_0 m^*c^2},
\end{equation}
where $N$ is the concentration of free charge carriers with effective mass $m^*$, $e$ is the elementary charge, $\varepsilon_0$ is the vacuum permittivity, $c$ the speed of light in vacuum, $n_0$ the refractive index of the undoped host material and $\lambda_0$ the vacuum wavelength for which we evaluate the refractive index. Increasing the free-carrier concentration $N$ will thus decrease the refractive index for 4H-SiC \cite{yoshida2012,sedighi2014}. This should allow for total internal reflection at shallow reflection angles.

\section{Sample materials}

The \mbox{n-i-n} planar waveguide samples used in this research are grown on an n$^{++}$ 4H-SiC substrate ($8\times10^{18}~\text{cm}^{-3}$). Using epitaxial chemical vapor deposition growth in which the nitrogen concentration for n-type doping could be controlled, first a $4~\mu$m layer of minimally doped ($10^{14}~\text{cm}^{-3}$) SiC is grown as the waveguide core. Next, a highly doped layer ($10^{19}~\text{cm}^{-3}$) of $2~\mu$m is deposited on top to serve as cladding. The material was then cleaved along the $[1\bar{1}00]$ direction and its orthogonal to yield several samples of 3.2 and 6.9~mm length with very smooth facets. 

The \mbox{p-i-n} planar waveguides were produced in a similar way, where aluminum is used for the highly p-doped layer. A minimally doped layer ($5\times10^{14}~\text{cm}^{-3}$) of $4.1~\mu$m thickness is deposited on top of a highly n-doped 4H-SiC substrate ($5\times10^{18}~\text{cm}^{-3}$), after which a layer with a highly p-doped layer ($2\times10^{19}~\text{cm}^{-3}$) of $2~\mu$m thickness is deposited on top of the intrinsic layer. Also this sample was cleaved along the $[1\bar{1}00]$ direction and its orthogonal, yielding samples of 2.8 and 5~mm length. Facets with rough edges were polished to improve the surface quality.

\begin{figure}[h]
	\centering
	\includegraphics[width=\linewidth]{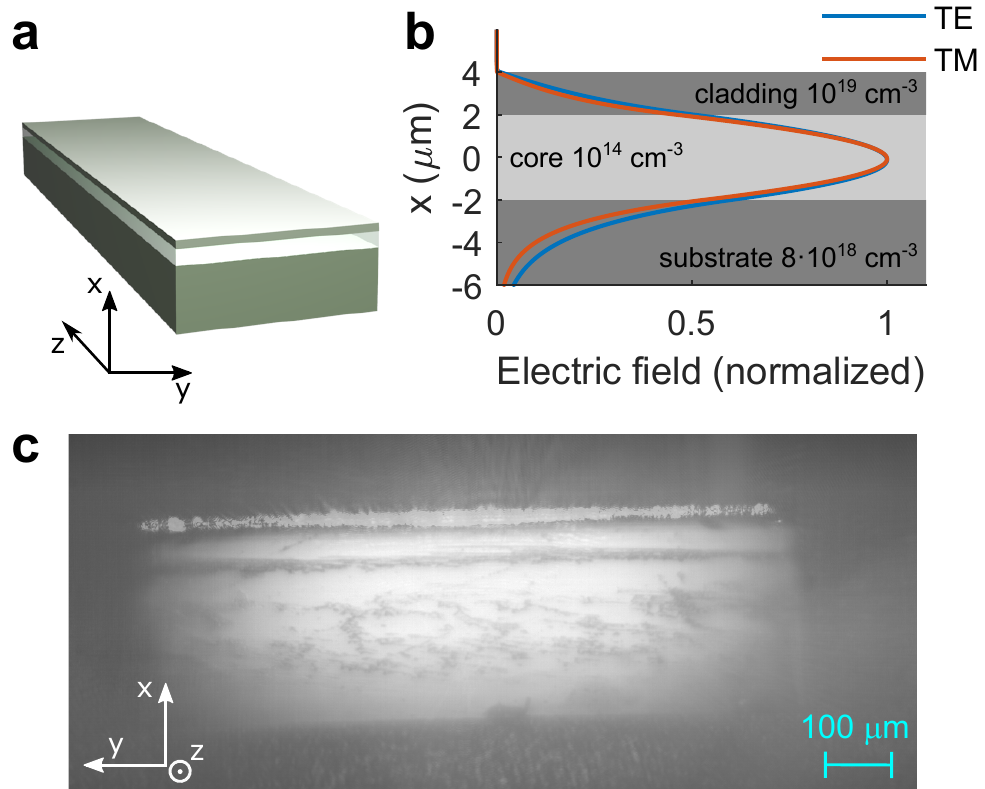}
	\caption{\textbf{Demonstration of waveguiding}. \textbf{a)} Schematic of a monolithic waveguide sample. On the substrate layer (bottom) an f = 4$~\mu$m core and a 2$~\mu$m cladding layer are grown. \textbf{b)} Predicted electric field distribution of the waveguide modes for TE and TM polarization, the distributions are independent of wavelength. \textbf{c)} Camera image of the end facet of a sample with an 800~nm laser coupled into the waveguide layer. The bright bar at the top results from transmission that is confined to a layer near the top of the sample. The remainder of the sample material is also well visible due to a white light source shining from the camera direction. Related video data is available online as Supplementary Material.}
	\label{fig:Fig_Alignment}
\end{figure}

According to Eq.~(\ref{eq:refractiveindex}) the step in index of refraction between the core and cladding is $\Delta n\approx3\times10^{-3}$ for 800~nm TE polarized light, where it is assumed that each n-dopant donates one electron to the conduction band or donates one hole to the valence band in the case of the $p^{++}$ doped cladding layer. With this knowledge we can predict the electric field distribution for the optical modes. The expected field distributions for TE and TM modes in the \mbox{n-i-n} junction are shown in Fig.~\ref{fig:Fig_Alignment}b \cite{ModeSolverOMS}. Due to the relatively low contrast in refractive index between the core and the cladding and substrate layer, wee see that a significant fraction of the electric field of the mode is outside the core layer. Interestingly, the shape of these modes is wavelength independent due to two competing effects. For increasing wavelengths, the reduction of round-trip phase for the optical field bouncing between the substrate and cladding layers is compensated by the increase in the range of total internal reflection angles \cite{hunsperger2009}. This is discussed in detail in Appendix~\ref{app:WLdeptMode}. 

Finally, we note that both the shallow n and p-dopants emit with near-bandgap energies, and mostly behave as contributors to the charge carrier plasma. Therefore, they have little optical interactions in the near-infrared spectral window of the color centers.

%-------------------------------------------------------------------
%--------------------- Experimental --------------------------------
%-------------------------------------------------------------------	

\section{Mode matching}

To investigate the waveguide functionality, we evaluate the transmission and confinement of a single-mode laser beam through the sample. For this purpose we used a 700-1000~nm tunable CW Ti:Sapphire laser, and tunable diode lasers for use around 1100 and 1300~nm. For initial testing we focused an 800~nm beam to a diameter of $4~\mu$m and aligned it to the waveguide layer using a stepper-motor controlled six-axis stage. Figure~\ref{fig:Fig_Alignment}c shows a camera image of the end facet of a sample for optimal mode coupling at the input side (related video data is available online as Supplementary Material). Along the top edge, a bright transmission band is visible, confirming the confinement of the optical field in the \mbox{n-i-n} core layer.

\begin{figure}[h]
	\centering
	\includegraphics[width=\linewidth]{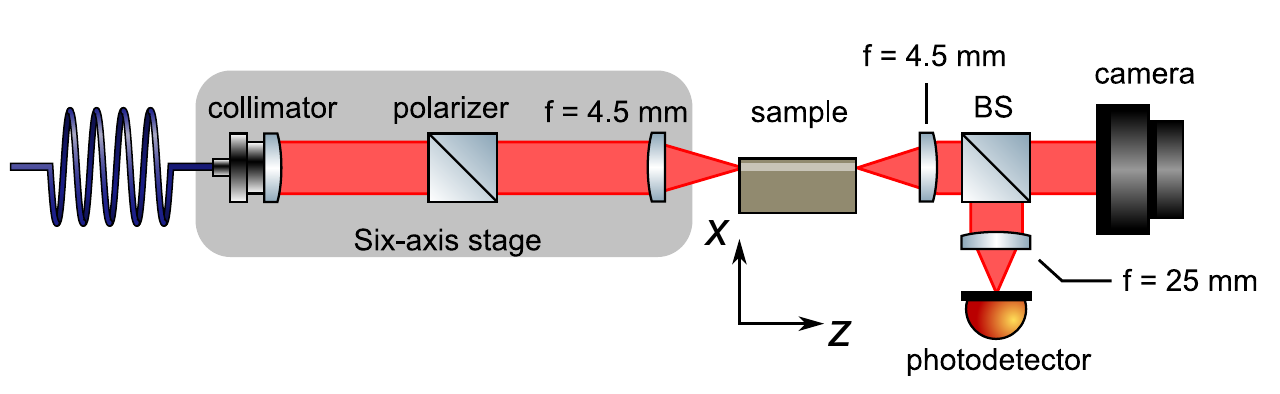}
	\caption{\textbf{Waveguide coupling setup.} A laser beam is fed to a six-axis stage by a single-mode fiber. After collimation and polarization it is focused onto a waveguide sample by an f = 4.5~mm aspheric lens. Half of the intensity of light exiting the sample is focused onto a CMOS camera, the other half is focused into a photodetector. The $x$- and $z$-axes are shown.}
	\label{fig:Fig_Setup}
\end{figure}

We further study the mode matching and coupling efficiency by measuring the transmission versus focusing-lens position (for the focal point near the input facet). Figure~\ref{fig:Fig_Setup} shows the experimental setup. A single-mode fiber attached to a six-axis stage feeds a wavelength-tunable laser to the setup. On the stage the laser beam with 1.2~mm diameter is polarized and focused by an aspheric lens with $4.5$~mm focal length. By scanning the position of the stage, the focal point of the laser can be scanned along the input facet of sample. The transmission is captured by a photodetector and a camera focused at the end facet of the sample. We repeat these scans for several wavelengths in order to study the wavelength dependence of the mode matching.

\begin{figure}[h!]
	\centering
	\includegraphics[width=\linewidth]{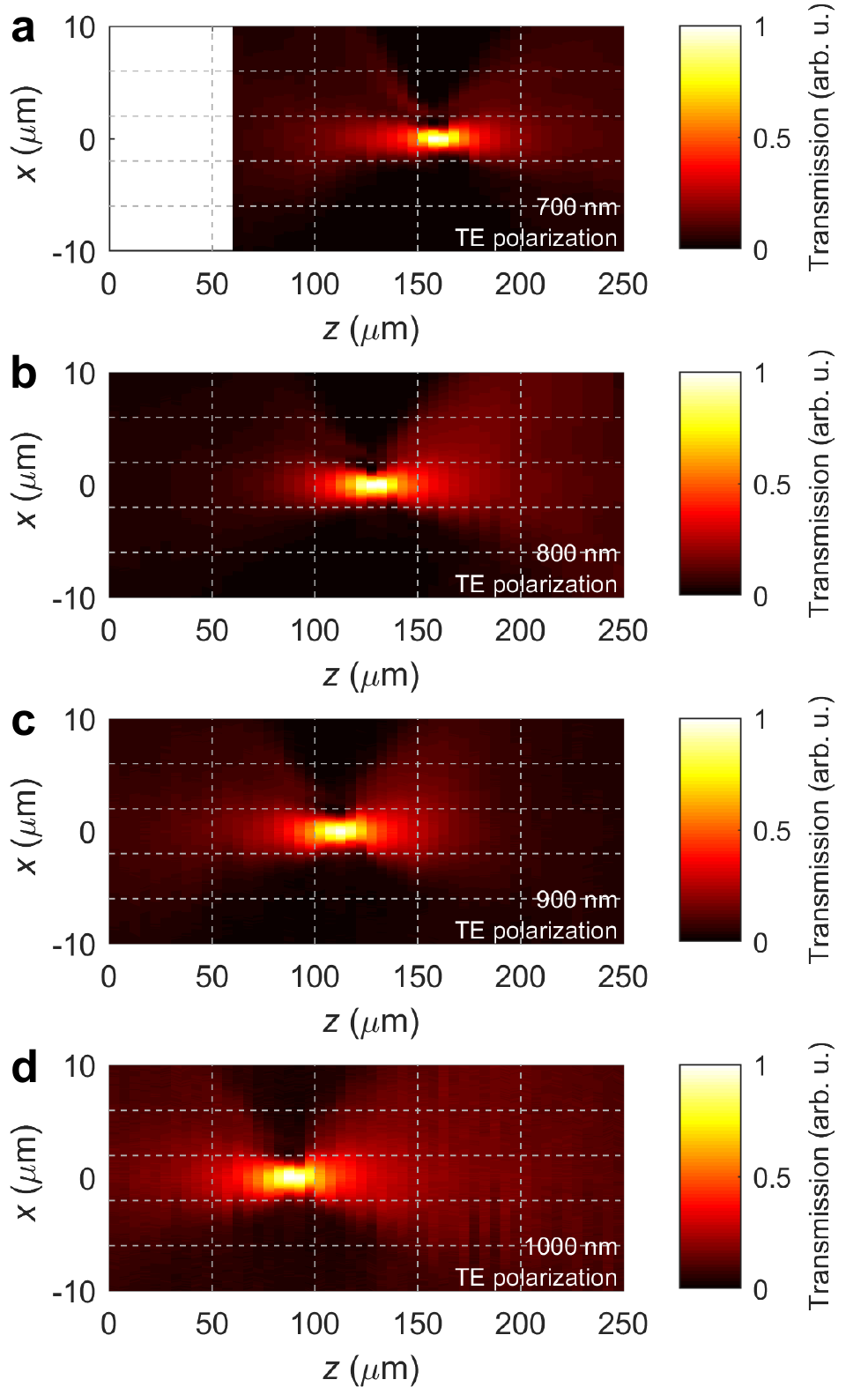}
	\caption{\textbf{Transmission for TE polarized light versus focusing-lens position for a \mbox{n-i-n} planar junction.} The focus of the coupling lens is scanned along the $z$-axis and along the sample layers ($x$-axis). \textbf{a-d)} Transmission for 700 - 1000~nm wavelength, respectively.}
	\label{fig:Fig_WLdept}
\end{figure}

Figure~\ref{fig:Fig_WLdept} presents the results for scanning along the optical axis ($z$) and sample layers ($x$) of the \mbox{n-i-n} junction for TE polarization, \textit{i.e.} polarization along the sample plane. Similar results for TM polarization, as well as results for longer wavelengths (up to 1290 nm), are presented in Appendix~\ref{app:ModeMatchTM-IR}. The results show a clear maximum in the amount of transmitted light, which reduces symmetrically according to a reduction in mode matching between the focus in a single-mode Gaussian beam and the waveguide mode, when shifting the lens in the $x$ and $z$ direction. This confirms that we couple in maximally when the input facet is fully in focus. We will further characterize this mode matching by determining the peak height and peak width (as full-width at half maximum, FWHM) of Gaussian fits of traces along the \mbox{$x$-axis} for data as in Fig.~\ref{fig:Fig_WLdept}. At the narrowest, the FWHM is $1.5~\mu$m, independent of wavelength. Note that the shift of the optimal coupling point along $z$ with wavelength matches with the wavelength dependence for the focal length of the used aspheric lens, as specified by the supplier \cite{C230TM_FocalShift}.

%-------------------------------------------------------------------
%---------------------- Figure WLdept  -----------------------------------
%-------------------------------------------------------------------

\begin{figure}[h!]
	\centering
	\includegraphics[width=\linewidth]{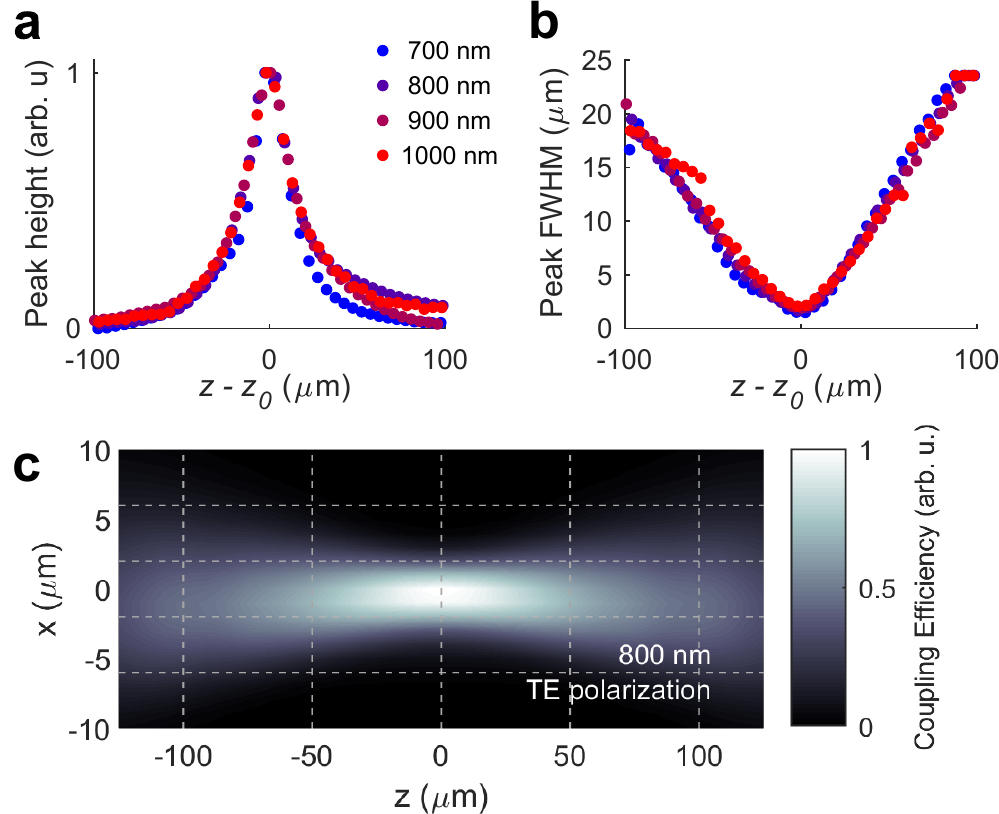}
	\caption{\textbf{Wavelength-independent coupling to a \mbox{n-i-n} planar junction for TE polarization.} \textbf{a)} Mode-matching peak height versus position along $z$. \textbf{b)} Mode-matching peak width along $x$ versus position along $z$. Both the peak width and peak height are obtained from fitting the transmission data in Fig.~\ref{fig:Fig_WLdept} to Gaussian curves along $x$. \textbf{c)} Theoretically predicted coupling efficiency for our experimental setup and waveguides (brighter means higher coupling efficiency).}
	\label{fig:Fig_PeakHeightWidth}
\end{figure}

Figure~\ref{fig:Fig_PeakHeightWidth}a,b present the mode-matching peak heights and FWHM for the four wavelengths used for measuring Fig.~\ref{fig:Fig_WLdept}a-d. Figure~\ref{fig:Fig_PeakHeightWidth}a shows the value of the transmission maximum for each position along $z$, relative to the overall transmission maximum at $z = z_0$. Figure~\ref{fig:Fig_PeakHeightWidth}b shows the FWHM. Both the peak width and height are unaffected by wavelength changes in this range, confirming that the mode distribution within the waveguide core is independent of wavelength. Figure~\ref{fig:Fig_PeakHeightWidth}c shows a theoretical prediction of the coupling efficiency for various beam positions at 800~nm. The coupling-efficiency profile matches well with the transmission profile from Fig.~\ref{fig:Fig_WLdept}b. This estimate was obtained by solving the overlap integral $\eta$ for optical powers \cite{chang2017} 
\begin{equation}
\eta = \frac{\abs{\int E_b^*E_m \text{d} A}}{\int \abs{E_b}^2 \text{d}A \int \abs{E_m}^2 \text{d} A},
\label{eq:overlap}
\end{equation} 
with $E_b$ the complex electric field of the laser beam and $E_m$ the electric field for the waveguide mode. These electric-field terms were determined based on the experimental-setup parameters and the predicted waveguide mode from Fig.~\ref{fig:Fig_Alignment}b, respectively.

%---------------------------------------------
% ------- p-i-n wavematching -------------------
%---------------------------------------------

\begin{figure}[h!]
	\centering
	\includegraphics[width=\linewidth]{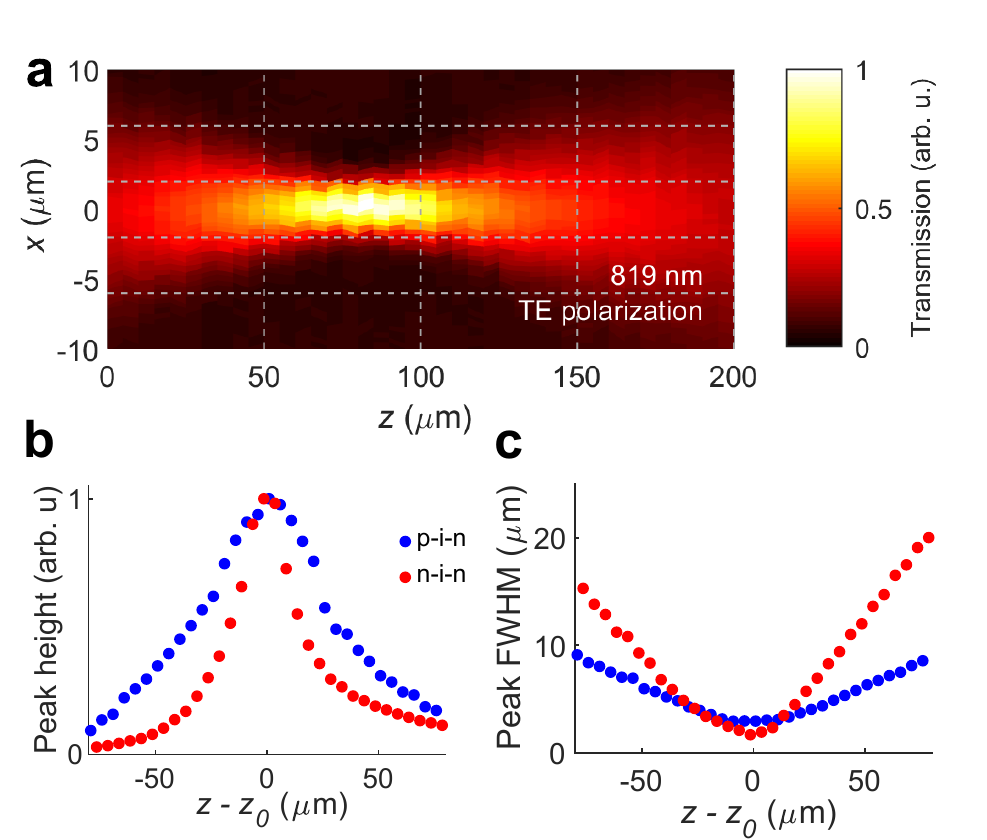}
	\caption{\textbf{Mode-coupling measurement and analysis for a \mbox{p-i-n} planar junction for 819~nm wavelength.} \textbf{a)} Dependence of transmission signal on the position of the focusing lens along the $x$ and $z$-axis. Note that the coordinate system has a different origin compared to Figs.~\ref{fig:Fig_WLdept} and \ref{fig:Fig_WLdept_TM}. \textbf{b)} Mode-matching peak height versus position along $z$. \textbf{c)} Mode-matching peak width (FWHM) along $x$ versus position along $z$. Both the peak width and peak height of the \mbox{p-i-n} junction are obtained from fitting the transmission data in a) to Gaussian curves along $x$, and are shown together with the $800$~nm data of the \mbox{n-i-n} junction in Fig.~\ref{fig:Fig_PeakHeightWidth}a,b.}
	\label{fig:Fig_pin}
\end{figure}

We repeated the mode-matching measurement of the \mbox{n-i-n} junction for the \mbox{p-i-n} junction with $819~\text{nm}$ light and TE polarization. The results are displayed in Fig.~\ref{fig:Fig_pin}, showing similar behavior to both the experimental results of the \mbox{n-i-n} junction and the theoretical prediction in Fig.~\ref{fig:Fig_PeakHeightWidth}c. Directly comparing results for the \mbox{p-i-n} junction to the results of the \mbox{n-i-n} junction (Fig.~\ref{fig:Fig_pin}b,c) shows that the mode-matching peak height and FWHM of the \mbox{p-i-n} junction depend less strongly on the $z$-position of the focusing-lens than for the \mbox{n-i-n} junction. This is because of the increased height of the mode (along $x$) in the \mbox{p-i-n} material. This results from the fact that the value $\Delta n$ is lower for this sample, due to a lower charge carrier concentration and the higher effective mass of holes compared to electrons \cite{Son2000}. Consequently, the  overlap integral in Eq.~(\ref{eq:overlap}) is less sensitive to a change in diameter of the $E_b$ field. This is also visible from the minimum peak FWHM, which is $2.9~\mu$m for the \mbox{p-i-n} junction (Fig.~\ref{fig:Fig_pin}c).

%---------------------------------------------
% ------- Propagation loss -------------------
%---------------------------------------------	

We obtained a rough indication for the propagation losses by comparing the transmission at optimal coupling for \mbox{n-i-n} and \mbox{p-i-n} samples of varied lengths ($3.2$ and $6.9$~mm, and $2.8$ and $5$~mm respectively). We found that for both TE and TM polarization the propagation losses must be below $14~\text{dB}/\text{cm}$ in the $700-1000~$nm wavelength range. We expect that a part of these losses is due to free-carrier (intraband) absorption in the highly doped layers around the core, where around $25\%$ of the electric field is present. Other mechanisms are surface scattering and losses due to imperfections in the setup, such incomplete collection of the light due to beam divergence inside the planar waveguide. To what extent each of these mechanisms contribute will be tested better in future work where these materials are processed into elongated one-dimensional waveguides.

\section{Conclusions}

In this research we have studied the feasibility of using strong doping of 4H-SiC for realizing cladding layers around a (near) intrinsic 4H-SiC optical waveguide core. This provides a path to waveguides in single-crystal high-purity 4H-SiC material. We fabricated monolithic planar waveguides that allow for broadband single-mode operation. We found as a rough upper bound for the propagation losses $14~\text{dB}/\text{cm}$,  which is low enough for on-chip applications. At a typical on-chip length scale of $100~\mu$m less than $3.2\%$ of the optical intensity will then be dissipated. As follow-up we propose to process and test actual one-dimensional photonic channels from these materials, and to engineer \mbox{n-i-n} or \mbox{p-i-n} junction waveguide with color-center defects in the waveguide core (such as divacancies or vanadium impurities \cite{koehl2011,zwier2015,widmann2019,saddow2011,simin2015,bosma2018,Spindlberger2019a,Wolfowicz2020}). This can demonstrate the true potential of these devices for quantum technologies with on-chip high-intensity optical driving of color centers.

\section{Supplementary Material}
See the supplementary material for video data that supports Fig.~\ref{fig:Fig_Alignment}c for different $z$-positions of the focusing-lens.

\begin{acknowledgements}
We thank R.~H.~van der Velde, F.~\v{S}imi\'c, R.~J.~M.~Julius, O.~V.~Zwier and D.~O'Shea for discussions and experimental contributions. We thank M.~de~Roosz, J.~G.~Holstein, T.~J.~Schouten, H.~H.~de Vries, F.~H.~ van der Velde and H.~Adema for technical support. We acknowledge support from the EU H2020 project QuanTELCO (all, grant No. 862721), the Knut and Alice Wallenberg Foundation (J.U.H and N.T.S., grant No. KAW 2018-0071), and the Swedish Research Council (N.T.S., grant No. 2016--04068 and J.U.H, grant No. 2020-05444).\newline
\end{acknowledgements}

\section*{Data Availability}
The data that support the findings of this study are available from the corresponding author upon reasonable request.

\section*{Author contributions}
The project was initiated by C.H.W. and T.B. SiC materials were grown and prepared by M.G., N.T.S. and J.U.H. Experiments were performed by T.B. and J.H. Data analysis was performed by T.B., J.H. and C.H.W. T.B., J.H. and C.H.W. had the lead on writing the manuscript.\newline

%%%%%%%%%%%%%%%%%%%%%%%%%%%%%%%%%
%%%%%%%%%% REFERENCES
%%%%%%%%%%%%%%%%%%%%%%%%%%%%%%%%%

\bibliography{Waveguide_bib}

%%%%%%%%%%%%%%%%%%%%%%%%%%%%%%%
%%%%%%%%% APPENDIX %%%%%%%%%%%%
%%%%%%%%%%%%%%%%%%%%%%%%%%%%%%%
\appendix

\section{}
\FloatBarrier

\subsection{Refractive index contrast}
\label{app:drude}

\paragraph*{Drude model}

We start from the following equation of motion for a driven mass-damper system \cite{fox2010}
\begin{equation}
\ddot{x} +  \gamma \dot{x} = -\frac{e}{m^*} E,
\end{equation}
where $x$ represents the displacement of the charge carrier, $\gamma$ is the damping constant, $e$ the electronic charge, $m^*$ the effective mass of the carrier and $E$ the applied electric field (associated with the incident light). For an oscillating field $E$ and a linear solution for $x$ we get
\begin{equation}
x=\frac{e/m^*E}{\omega^2+i\gamma\omega}.
\end{equation}
The contribution to the polarization from free carriers that results from the doping is then
\begin{equation}
P_\text{doping} = - \frac{Ne^2E}{m^*(\omega^2+i\gamma\omega)},
\end{equation}
with $N$ the number density of additional nearly-free charge carriers in the material. This contribution to the polarization perturbs the total electric displacement as
\begin{equation}
D = \varepsilon_0 E + P_\text{material} + P_\text{doping},
\end{equation}
where $\varepsilon_0$ is the permittivity of free space and $P_\text{material}$ the response of the undoped material to the applied field $E$. The electric displacement $D$ of the pure material can also be written using the dielectric constant. We get
\begin{align}
D&= \varepsilon_0\varepsilon_\text{opt}E + P_\text{doping},\\
&= \left(\varepsilon_0\varepsilon_\text{opt} -\frac{Ne^2}{m^*(\omega^2+i\gamma\omega)}\right)E,	
\end{align}
where $\varepsilon_\text{opt}=n_0^2\approx6.8$ \cite{wang2013} is the dielectric constant of undoped 4H-SiC. The dielectric constant of the doped layer is then
\begin{equation}\label{eq:refractiveindexApp}
\varepsilon_r = \varepsilon_\text{opt}  -\frac{Ne^2}{\varepsilon_0m^*(\omega^2+i\gamma\omega)} = n_\text{doped}^2,
\end{equation}
where $n_\text{doped}$ is the refractive index of the doped layer.

It should be noted that Eq.~(\ref{eq:refractiveindexApp}) is the same for positive and negative charge carriers. Either of these will lower the dielectric constant of the material with respect to undoped material. We assume light damping for the plasma, so we can write
\begin{equation} \label{eq:eps_rApp}
n_\text{doped}^2 = n_0^2 \left(1 - \frac{\omega_p^2}{\omega^2}\right),
\end{equation}
where
\begin{equation} \label{eq:plasmeFrequency}
\omega_p=\sqrt{\frac{Ne^2}{m^*\varepsilon_0\varepsilon_\text{opt}}}
\end{equation}
is the plasma frequency. 

\paragraph*{Contrast in refractive index}

The contrast in refractive between doped and an undoped layers of 4H-SiC can now be calculated as function of doping level $N$ (assuming that each dopant adds one nearly-free charge carrier). It is worth considering that there will always be some degree of unintentional doping. For our material, this level is around $10^{14}\si{~cm^{-3}}$. The contrast will be defined as
\begin{equation} \label{eq:DeltaNApp}
\Delta n = n_\text{doped} - n_0.
\end{equation}
Substituing Eq.~(\ref{eq:eps_rApp}) into Eq.~(\ref{eq:DeltaNApp}) we get
\begin{equation}
\Delta n = n_0 \left(\sqrt{1 - \frac{\omega_p^2}{\omega^2}} - 1 \right),
\end{equation}
which, after applying the Taylor series expansion, becomes 
\begin{equation}
\Delta n \approx -n_0 \left( \frac{\omega_p^2}{2 \omega^2} \right).
\end{equation}
After substituting Eq.~(\ref{eq:plasmeFrequency}) and using $\varepsilon_\text{opt}=n_0^2$ again, we arrive at Eq.~(\ref{eq:refractiveindex}).

\begin{figure}[h]
	\centering
	\includegraphics[width=\linewidth]{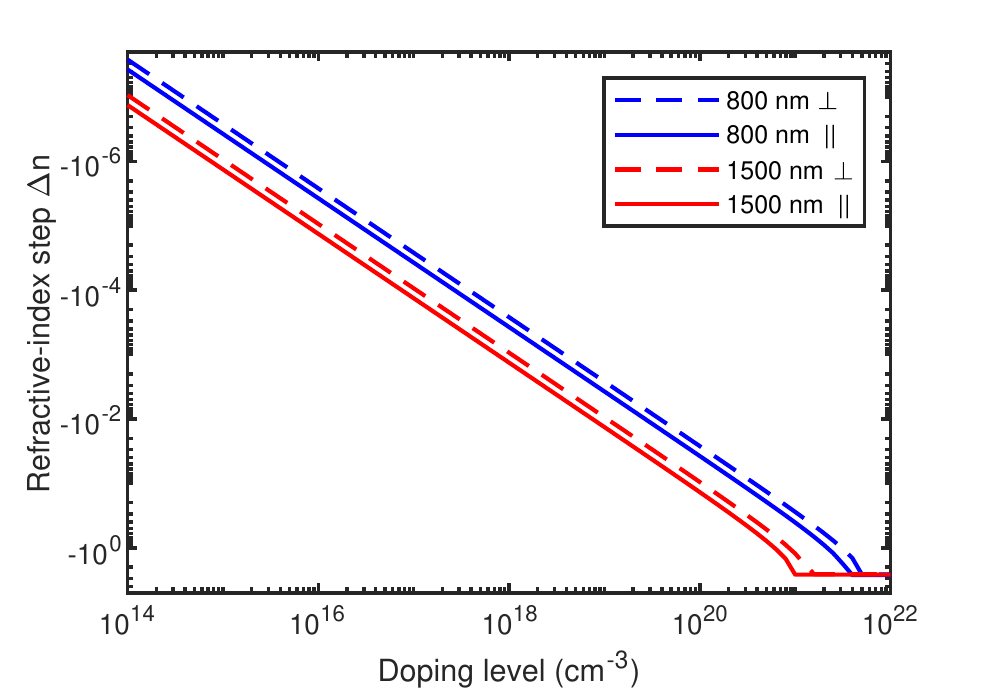}
	\caption{\textbf{4H-SiC refractive-index contrast versus doping level.} The step in index of refraction $\Delta n$ is plotted versus doping level for 800 and 1500~nm light and for linear polarization parallel $\parallel$ to the crystal c-axis (TM) and orthogonal $\perp$ to the c-axis (TE), for the geometry that the c-axis is orthogonal to the waveguide plane.}
	\label{fig:contrastvsdoping}
\end{figure}

Figure \ref{fig:contrastvsdoping} shows the dependence of contrast in refractive index $\Delta n$ on doping level. At a doping level of $10^{19}\si{~cm^{-3}}$, the contrast $\Delta n$ reaches a value of $3\times10^{-3}$, which results in total internal reflection at angles above $\ang{87.4}$ for light propagating from an undoped to a doped region of 4H-SiC.

For doping levels above $10^{21}\si{~cm^{-3}}$, the contrast reaches a saturation as the real part of the refractive index becomes zero. At this point the plasma frequency from Eq.~(\ref{eq:plasmeFrequency}) exceeds the frequency of the incident light and the material exhibits metallic reflection.

\subsection{Wavelength independence for TE and TM waveguide modes}
\label{app:WLdeptMode}

In this section we present an analysis that clarifies why the shape of the optical mode in the waveguide is nearly independent on wavelength of the light, for all cases where the associated optical frequency is well above the plasma frequency of the free carrier system, building on Refs.~\onlinecite{saleh2007,hunsperger2009}.
In order to model the electric field distribution inside a planar waveguide, we have to consider the wave nature of light. For certain angles of reflection at the core-substrate and core-cladding interfaces the phase of the wave yields constructive interference with itself after the two reflections. In such cases the total field can be described by two particular plane waves that interfere with each other to generate a field distribution that is homogeneous along the propagation direction. These particular fields define the waveguide modes.

\paragraph*{Reflection angle $\theta_m$}

\begin{figure}[h!]
	\centering
	\includegraphics[width=\linewidth]{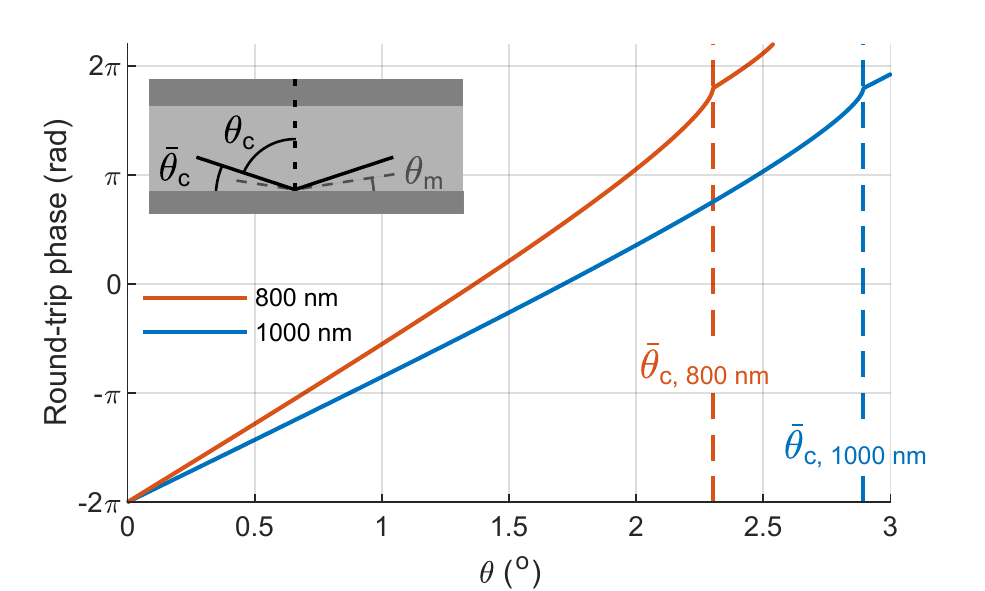}
	\caption{\textbf{Round-trip phase versus bounce angle for TE polarization.} Phase accumulation upon two reflections versus reflection angle between the $k$-vector and the core/cladding interface, for 800 and 1000~nm wavelengths. The red and blue-dashed lines represent the complement of the critical angle for total internal reflection $\bar{\theta}_\text{c}$ on the core-substrate interface of our sample for 800 and 1000~nm, respectively. The inset shows how this complement angle is defined. Below $\bar{\theta}_\text{c}$ total internal reflection occurs.}
	\label{fig:Fig_RoundTripPhase}
\end{figure}

The reflection angle $\theta_m$ for the mode $m$ (where $m=0$ is the lowest order mode) can be found using the mentioned self-consistency condition where the wave should reproduce itself every second reflection. For a dielectric waveguide, this condition is                                                                            
\begin{equation}
4\pi \frac{n_\text{core}d}{\lambda_0}\sin{\theta_m} + \varphi_\text{clad} + \varphi_\text{sub} = 2\pi m,
\end{equation}
where $\lambda_0$ is the vacuum wavelength, $n_\text{core}$ the refractive index for the waveguide core, $d$ the thickness of the core, and $\varphi_\text{clad(sub)}$ is the phase that is accumulated upon reflection from the cladding (substrate) interface. The latter be found by solving the boundary value problem for the electromagnetic field at the core-cladding and core-substrate interfaces. For TE polarization it becomes
\begin{equation}\label{eq:RoundTripPhase}
\varphi_\text{clad(sub)} =  \tan^{-1}\left(\sqrt{\frac{\cos^{2}{{\theta}_\text{c,clad(sub)}}}{\sin^2{\theta_m}}-1}\right),
\end{equation}
with ${{\theta}_\text{c,clad(sub)}}$ the critical angle for total internal reflection on the cladding (substrate). Figure~\ref{fig:Fig_RoundTripPhase} shows the accumulated phase as a function of reflection angle $\theta$, for 800~nm and 1000~nm light for the waveguide structures as mentioned in the main text. The total round-trip phase will not pass $2\pi$ before the critical angle for total internal reflection on the core-substrate interface $\bar{\theta}_\text{c,sub}$ is reached. Therefore, this waveguide will only support the fundamental waveguide mode.

In fact, for the doping concentrations mentioned in the main text, this will be a single-mode waveguide for all wavelengths \footnote{That is: for all frequencies below the bandgap and far enough above the plasma frequency}. This is because the complement of the critical angle  $\bar{\theta}_\text{c}$ scales linearly with wavelength in the small-angle approximation for $\bar{\theta}_\text{c}$ and binomial approximation for $\Delta n$. The critical angle becomes
\begin{align*}
\cos\bar{\theta}_c &= \frac{n_\text{core} + \Delta n}{n_\text{core}},\\
1 - \frac{\bar{\theta}_c^2}{2} &\approx 1 - \frac{\omega_p^2}{2  n_\text{core}^2 c^2} \lambda_0^2,\\
\bar{\theta}_c & \approx \frac{\omega_p}{ n_\text{core} c}\lambda_0.
\end{align*}
Therefore, the slower phase accumulation (Eq.~(\ref{eq:RoundTripPhase})) at larger wavelengths is near-perfectly compensated by an increased $\bar{\theta}_c$.

\subsection{Additional results for mode-matching studies}
\label{app:ModeMatchTM-IR}

\begin{figure}[h!]
	\centering
	\includegraphics[width=\linewidth]{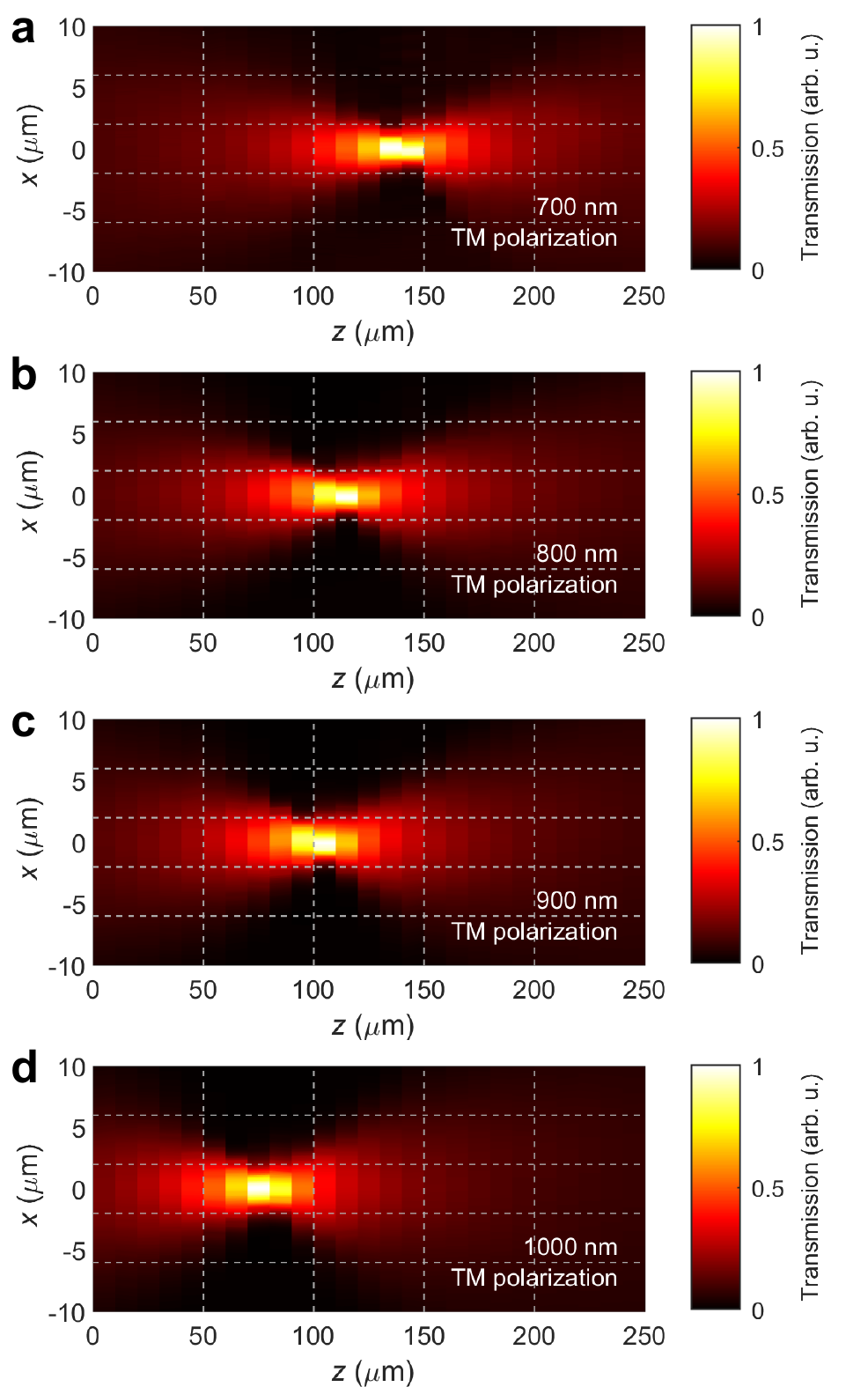}
	\caption{\textbf{Transmission versus focusing-lens position for a \mbox{n-i-n} planar junction} The coupling lens is scanned along the $z$-axis and along the sample layers ($x$-axis), for the focus near the waveguide entrance facet (the exact focal point depends on wavelength). \textbf{a-d)} Transmission for 700 - 1000~nm wavelength, respectively.}
	\label{fig:Fig_WLdept_TM}
\end{figure}

\begin{figure}[h]
	\centering
	\includegraphics[width=0.7\linewidth]{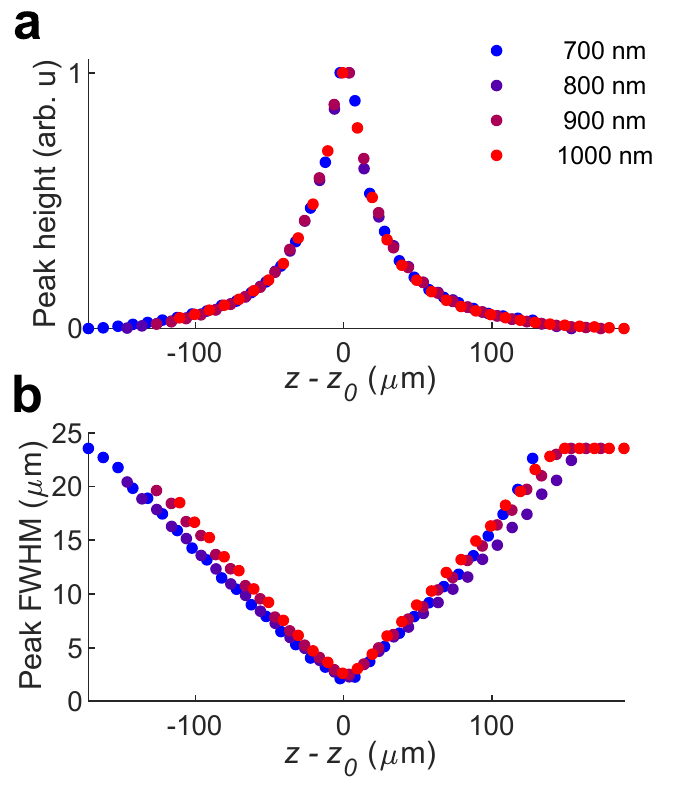}
	\caption{\textbf{Wavelength-independent coupling for TM polarization to a \mbox{n-i-n} planar junction.} \textbf{a)} Mode-matching peak height versus position along $z$. \textbf{b)} Mode-matching peak width along $x$ versus position along $z$. Both the peak width and peak height are obtained from fitting the transmission data in Fig.~\ref{fig:Fig_WLdept_TM} to Gaussian curves along $x$.}
	\label{fig:Fig_PeakHeightWidth_TM}
\end{figure}

\begin{figure}[h!]
	\centering
	\includegraphics[width=\linewidth]{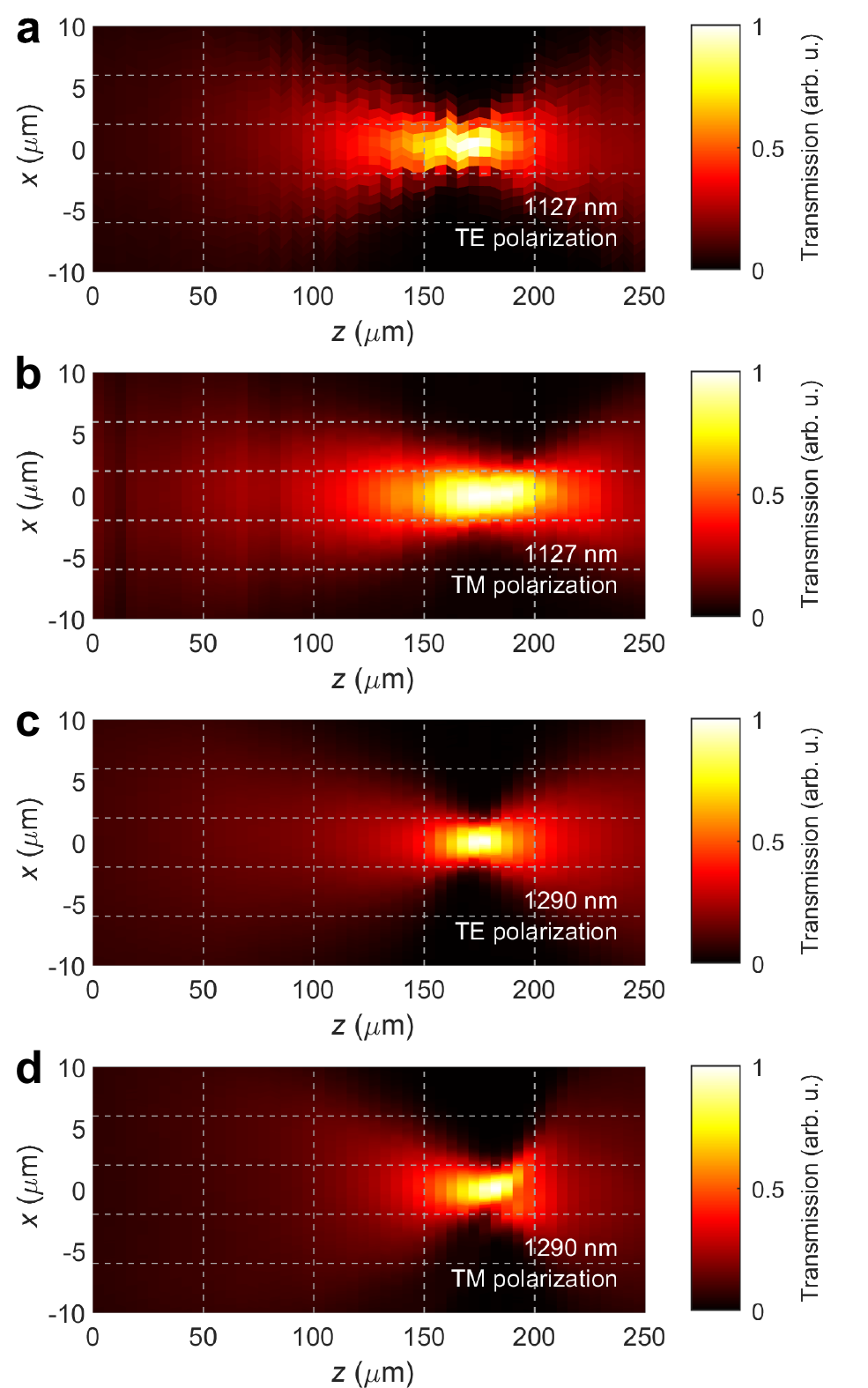}
	\caption{\textbf{Transmission versus focusing-lens position at 1127 and 1290~nm for a \mbox{n-i-n} junction.} The focus of the coupling lens is scanned along the $z$-axis and along the sample layers ($x$-axis). \textbf{a-b)} Transmission at 1127~nm for TE and TM polarization, respectively. \textbf{c-d)} Transmission at 1290~nm for TE and TM polarization, respectively. Note that the coordinate system has a different origin compared to Figs.~\ref{fig:Fig_WLdept}, \ref{fig:Fig_pin} and \ref{fig:Fig_WLdept_TM}.}
	\label{fig:Fig_WLdept_TM_1300}
\end{figure}

Figure~\ref{fig:Fig_WLdept_TM} presents experimental results for TM polarization and wavelengths in the range 700 to 1000~nm. An analysis of these mode-matching results is presented in Fig.~\ref{fig:Fig_PeakHeightWidth_TM}. Figure~\ref{fig:Fig_WLdept_TM_1300} presents experimental results for both TE and TM polarization, for the wavelengths 1127 and 1290~nm. 

\FloatBarrier

\end{document}